\documentclass[%
 reprint,
 superscriptaddress,
 amsmath,
 amssymb,
 aps,
 pra,
 10pt
]{revtex4-2}

\usepackage{physics}
\usepackage[dvipsnames]{xcolor} 
\usepackage[hidelinks]{hyperref} 
\usepackage{hyperref}
\usepackage{mathtools}
\usepackage{verbatim}
\usepackage{graphicx}
\usepackage{dcolumn}
\usepackage{bm}
\usepackage{ulem}

\usepackage[group-minimum-digits=0,table-figures-decimal=0,table-number-alignment=center]{siunitx}
\usepackage[indent=10pt, skip=2pt plus1pt]{parskip} 
\usepackage{makeidx}
\usepackage{lipsum} 

\hypersetup{
    colorlinks=true,
    breaklinks=true,
    urlcolor=black,
    linkcolor=NavyBlue,
    citecolor=NavyBlue,
    bookmarksopen=true,
    pdftoolbar=false,
    pdfmenubar=false,
}

\newcommand{\ie}{i.e.}

\newcommand{\hc}{\text{h.c.}} 

\newcommand{\nbar}{\bar{n}}
\newcommand{\adag}{\hat{a}^\dagger}
\newcommand{\adagsq}{\hat{a}^{\dagger 2}}
\newcommand{\hata}{\hat{a}}

\makeindex

\begin{document}

\title{Two-photon driven Kerr quantum oscillator with multiple spectral degeneracies}
\author{Diego Ruiz}
\email{diego.ruiz@alice-bob.com}%
\affiliation{Alice \& Bob, 53 Bd du Général Martial Valin, 75015 Paris, France}
\affiliation{Laboratoire de Physique de l'École Normale Supérieure, École Normale Supérieure, Centre Automatique et Systèmes, Mines Paris, Université PSL, Sorbonne Université, CNRS, Inria, 75005 Paris}
\author{Ronan Gautier}%
\affiliation{Laboratoire de Physique de l'École Normale Supérieure, École Normale Supérieure, Centre Automatique et Systèmes, Mines Paris, Université PSL, Sorbonne Université, CNRS, Inria, 75005 Paris}
\author{Jérémie Guillaud}%
\affiliation{Alice \& Bob, 53 Bd du Général Martial Valin, 75015 Paris, France}
\author{Mazyar Mirrahimi}%
\affiliation{Laboratoire de Physique de l'École Normale Supérieure, École Normale Supérieure, Centre Automatique et Systèmes, Mines Paris, Université PSL, Sorbonne Université, CNRS, Inria, 75005 Paris}

\date{\today}

\begin{abstract}
Kerr nonlinear oscillators driven by a two-photon process are promising systems to encode quantum information and to ensure a hardware-efficient scaling towards fault-tolerant quantum computation. In this paper, we show that an extra control parameter, the detuning of the two-photon drive with respect to the oscillator resonance, plays a crucial role in the properties of the defined qubit. At specific values of this detuning, we benefit from strong symmetries in the system, leading to multiple degeneracies in the spectrum of the effective confinement Hamiltonian. Overall, these degeneracies lead to a stronger suppression of bit-flip errors. We also study the combination of such Hamiltonian confinement with colored dissipation to suppress leakage outside of the bosonic code space. We show that the additional degeneracies allow us to perform fast and high-fidelity gates while preserving a strong suppression of bit-flip errors.
\end{abstract}


\maketitle


\section{\label{sec:level1}Introduction}

Superconducting quantum circuits are one of the most promising and advanced platforms for the realization of quantum processors, with the capability of solving intractable problems for classical computers~\cite{Google_Supremacy}. An artificial atom, e.g. the transmon qubit~\cite{Koch2007}, is typically used to encode the information. Because of its high error rates compared to what is required to realize reliable useful algorithms, a large qubit overhead is needed for quantum error correction and fault-tolerance~\cite{Shor1996,Fowler2012}. The use of more advanced  qubits with some level of intrinsic protection against decoherence could potentially lead to an important reduction in such hardware complexity.

In this regard, encoding the qubit information in fancy states of harmonic oscillators has recently attracted an increasing interest. Indeed, benefiting from the infinite dimensional Hilbert space of  a bosonic mode, we can delocalize the quantum information in different parts of the harmonic oscillators phase space, and thus provide a first layer of protection. GKP states~\cite{Gottesman2001}, by encoding the information in a two-dimensional grid of infinitely squeezed states, could be concatenated with a smaller distance surface code~\cite{Fukui-PRL-2017,Fukui-PRX-2018,Vuillot-PRA-2019,Noh-PRA-2020}, compared to the ones needed for transmon qubits. Cat qubits encode the quantum information in superpositions of coherent states~\cite{Cochrane1999,Leghtas2013,Mirrahimi2014}. The bit-flip errors, which correspond to a transition from one coherent state to the other, are suppressed exponentially with their phase space separation, while the phase-flip errors only increase linearly~\cite{Lescanne2020}. This tunable noise bias leads to important reductions of hardware overhead for quantum error correction~\cite{Aliferis2008,tuckett2019tailoring,combes2022homodyne}. The recent proposal of bias-preserving gates~\cite{Guillaud2019,Puri2020} has paved the way towards hardware-efficient fault-tolerant and universal quantum processors based on such qubits~\cite{Guillaud2020,Chamberland2022,Darmawan2021}.

Two approaches have been considered so far to confine the cat states code space. One approach uses an engineered two-photon dissipation, whose only two steady states are the cat states~\cite{Mirrahimi2014,azouit2015convergence,Leghtas2015,Touzard2018,puri2019stabilized}. The other approach is to use a Hamiltonian confinement, combining a squeezing drive with Kerr non-linearity~\cite{Puri2017,Grimm2020,kwon2022autonomous,goto2019quantum,kanao2021high,goto2016universal,kanao2022quantum,chono2022two}. The cat states are the degenerate ground states of this Kerr Hamiltonian, protected by an energy gap proportional to the Kerr non-linearity.

In the dissipative approach, if the two-photon dissipation rate is faster than the typical error rates, any leakage outside  the code space is countered by the dynamical stabilization induced by the dissipative mechanism, without the encoded information leaking throughout this channel. This can be seen as autonomous error correction. The  resulting exponential suppression of bit-flip errors has been experimentally demonstrated~\cite{Lescanne2020} and is expected to reach macroscopic time-scales in forthcoming experiments~\cite{Berdou2022,gravina2022critical,wang2016schrodinger}. However, the bias-preserving gates are typically slow and making them faster than the dissipation timescale leads to an important increase of phase-flip errors~\cite{Guillaud2019}. 

In the Kerr Hamiltonian approach, the Hamiltonian dynamics makes it possible to perform gates in fast timescales of the order of the Kerr non-linearity, thanks to the adiabatic theorem and the design of super-adiabatic pulses~\cite{Xu2021}. However, while the energy gap confines the cat-qubit subspace, the leakage induced by mechanisms such as thermal excitation or photon dephasing is not countered by any stabilization process. They could hence lead to significant contributions in terms of bit-flip errors. Consequently, bit-flip errors are suppressed at a slower rate when increasing the number of encoding photons~\cite{Putterman2022,Gautier2022,Frattini-2022}. Finally, more recent studies~\cite{Gautier2022} have shown that it is not straightforward to combine the dissipative scheme with the Kerr Hamiltonian because there is no sweet spot between the Kerr non-linearity being too small to benefit from faster gates, or too large which compromises the exponential suppression of bit-flip errors. 

Recently, the detuning of the squeezing drive with respect to the Kerr resonator frequency has been used as an additional control knob to enhance the cat qubit performance~\cite{Frattini-2022,gravina2022critical}. In this paper, following the new bistability regimes discussed in~\cite{Roberts2019}, we demonstrate that the preservation of locality is significantly improved for specific values of this detuning. Indeed, we will prove that by appropriately choosing the parameters of such a Hamiltonian in the aforementioned bistable regimes, one can make sure that not only the two ground states of the effective confinement Hamiltonian are perfectly degenerate, but also that excited levels come in perfectly degenerate pairs. This degeneracy was also discussed in~\cite{marthaler2007,zhang2017} and used for preparing quasi-energy states of the driven system.  This degeneracy implies that leakage to such excited levels does not lead to non-local excursions in phase space. The information remaining localized in the phase space, it is hence possible to combine the Hamiltonian confinement with an appropriate weak dissipative mechanism to counter the leakage. The defined qubit would therefore benefit from reasonably low-rate bit-flip errors, while the strong Hamiltonian confinement enables fast bias-preserving gates.  

Section \ref{sec:level2} recalls why the Kerr Hamiltonian bit-flip probability is particularly sensitive to thermal excitations and dephasing and how this can be related to the Hamiltonian spectrum. It then introduces the new detuned Kerr Hamiltonian. We analyze its spectrum and explain why it provides a better protection against incoherent perturbations. Section \ref{sec:level4} demonstrates the gain on phase space locality compared to regular Kerr cats with numerical master equation simulations. Section \ref{sec:colored} presents the combination with a colored dissipation scheme to limit the leakage outside of the code space. Section \ref{sec:level6} presents the bias-preserving gates on this new encoding.

\section{\label{sec:level2}Squeezed Kerr oscillator and multiple degenerate bistable regimes}

Cat qubits are encoded in a quantum harmonic oscillator within the two-dimensional subspace defined by two coherent states of opposite phase. The codespace can be defined as follows~\cite{Cochrane1999,Mirrahimi2014},
\begin{equation}
    \begin{split}
        \ket{0}_c=\frac{1}{\sqrt{2}}(\ket{+}_c+\ket{-}_c) = \ket{+\alpha}+\mathcal{O} (e^{-2\abs{\alpha}^2}) \\
        \ket{1}_c=\frac{1}{\sqrt{2}}(\ket{+}_c-\ket{-}_c) = \ket{-\alpha}+\mathcal{O}(e^{-2\abs{\alpha}^2})
        \label{eq:CatEncoding}
    \end{split}
\end{equation}
where $\ket{\pm}_c=\mathcal{N}_{\pm}(\ket{\alpha}\pm\ket{-\alpha})$ with $\ket{\alpha}$ a coherent state of complex amplitude $\alpha$ and $\mathcal{N}_{\pm}=[2 (1\pm e^{-2 \abs{\alpha}^2})]^{-1/2}$ is a normalization constant. The states $\ket{+}_c$ and $\ket{-}_c$ are respectively called even and odd cat state and are also denoted $\ket{\mathcal{C_\alpha^\pm}}$.

\begin{figure*}[t!]
    \centering
    \includegraphics[width=\textwidth]{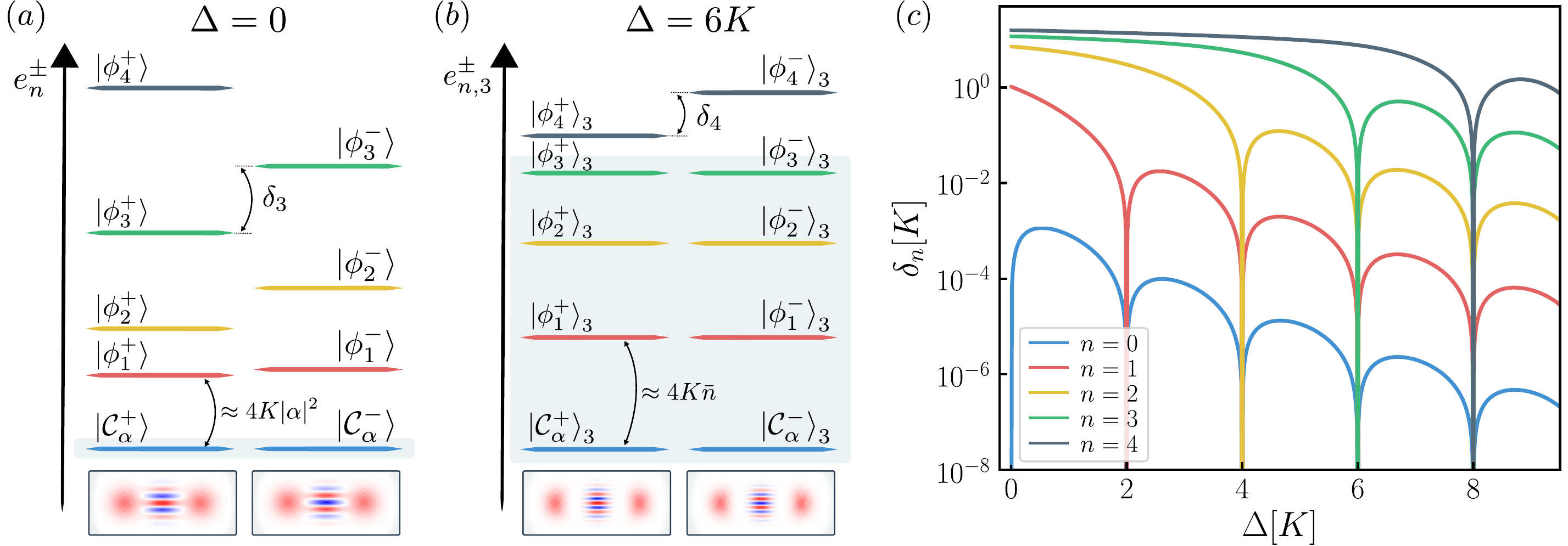}
    \vspace{-0.5cm}
    \caption{\label{fig:arche}
    (a) Energy levels of the Kerr Hamiltonian for $|\alpha|^2=4$, separated into two sets of even and odd photon-number parity states. The two ground states encode the cat qubits, protected by an energy gap $4 K |\alpha|^2$ from coherent perturbations. (b) Energy levels of the detuned Kerr Hamiltonian for $\Delta = 6K$ and $|\alpha|^2=\epsilon_2/K=4$. The first 4 pairs of eigenstates are exactly degenerate. The ground states are no longer cat states but are slightly deformed. Their mean photon number is given by $\bar n>|\alpha|^2$ for $\Delta>0$. (c) Spacing $\delta_n$ between the energy levels $\ket{\phi_n^+}$ and $\ket{\phi_n^-}$ for $|\alpha|^2=\epsilon_2/K=4$ as a function of the detuning $\Delta$. Both the energy level spacings and the detuning are in units of Kerr strength $K$. When $\Delta = 2mK$, the $2m+2$ first energy levels come in $m+1$ pairs of exactly degenerate states.
    }
\end{figure*}

In an appropriate rotating frame, the Hamiltonian of the two-photon squeezed Kerr oscillator reads~\cite{Puri2017}
\begin{equation}\label{eq:Kerr}
    \begin{split}
        \hat{H} &= K\hat{a}^{\dagger 2}\hat{a}^{ 2} + \epsilon_2 \hat{a}^{\dagger 2} + \epsilon_2^* \hat{a}^{2} \\
        &= K(\hat{a}^{\dagger 2} - \alpha^{*2})(\hat{a}^{ 2} - \alpha^2)
    \end{split}
\end{equation}
where $K$ denotes the Kerr non-linearity of the oscillator, $\epsilon_2$ the two-photon drive and $\pm \alpha\ (\alpha = \sqrt{-\epsilon_2/K})$ are the amplitudes of the two coherent states that are also ground states of the Kerr Hamiltonian. In Fig.~\ref{fig:arche}(a) we plot the eigenenergies of Hamiltonian~\eqref{eq:Kerr}. The two cat states are separated from the rest of the spectrum by an energy gap approximately given by $4 K \abs{\alpha}^2 $. We have grouped the higher eigenstates by their photon-number parity and denote them $\ket{\phi_n^\pm}$ with $e^\pm_n$ their respective energies. According to the quantum adiabatic theorem, the energy gap protects the ground subspace against weak and slow Hamiltonian perturbations.

However, incoherent perturbations due for instance to thermal excitations or Markovian photon dephasing leak the cat states to higher excited states. For instance, thermal excitations leak $\ket{\alpha}$ into $\ket{\alpha,1} = \hat{D}(\alpha) \ket{1}$ up to first order, where $\hat{D}$ is the displacement operator, and $\ket{\alpha,1}$ has a non-zero overlap with all eigenstates $\{\ket{\phi_n^\pm}\}_{n \ge 1}$. These higher eigenstates come in nearly degenerate pairs but their non-degeneracy is enough to generate a rather fast dephasing between odd and even photon number parity subspaces. In the phase space picture, such dephasing can be seen as non-local excursions between left and right half planes. This ultimately breaks down the protection of the Kerr cat qubit against bit-flip errors.

In order to capture the above non-local excursions in phase space, let us consider the average value of the observable $\hat S=\text{sign}(\hat X)=\text{sign}(\hat a+\hat a^\dag)$,
\begin{equation}
    \bra{\psi}\hat S \ket{\psi}=\int_0^\infty |\psi(x)|^2dx-\int_{-\infty}^0 |\psi(x)|^2 dx. 
\end{equation}
Following the calculations of~\cite{Gautier2022}, starting from one of the cat-qubit computational states, the dynamics of $|\langle \hat S\rangle|$ is well approximated by $\exp(-\Gamma t)$, with
\begin{equation}
    \begin{split}
        \Gamma &= \kappa_1|\alpha|^2e^{-4|\alpha|^2} + \kappa_le^{-2|\alpha|^2} \\
        & \quad +\kappa_l\sum_{n>0} \lambda_n \left[ 1- \frac{\text{sin}(\delta_n/\kappa_{\text{conf}}
        )}{\delta_n/\kappa_{\text{conf}}} \right].
        \label{eq:formuleRonan}
    \end{split}
\end{equation}
Here, $\kappa_l$ is the rate of leakage outside of the manifold span$\{\ket{\mathcal{C}_\alpha^+},\ket{\mathcal{C}_\alpha^-}\}$. Typically, taking a thermal excitation rate $n_{\text{th}}\kappa_1$ ($\kappa_1$ denoting the relaxation rate of the harmonic oscillator), and possibly a Markovian photon dephasing rate of $\kappa_\phi$, the leakage rate is given by $\kappa_l=n_{\text{th}}\kappa_1+|\alpha|^2\kappa_\phi$. Furthermore, $\lambda_n$ denotes the average overlap between $\ket{\alpha,1}$ and the eigenstates of the driven Kerr Hamiltonian $\ket{\phi_n^\pm}$. It reads $\lambda_n=\sum_{\pm}\abs{\braket{\alpha,1}{\phi_n^\pm}}^2/2$. Finally, $\delta_n$ is the energy level spacing between $\ket{\phi_n^-}$ and $\ket{\phi_n^+}$, $\delta_n = e^-_n - e^+_n$, and $\kappa_\text{conf}$ the dissipative confinement rate of the cat manifold. In the absence of non-linear dissipative mechanisms~\cite{Mirrahimi2014} or colored dissipation~\cite{Putterman2022}, $\kappa_\text{conf}$ is simply given by $\kappa_1$, the energy relaxation rate of the harmonic oscillator.

The last term in Eq.~\eqref{eq:formuleRonan} highlights the contribution of each pair of nearly degenerate excited eigenstates to the non-local excursion, leading in term to bit-flip errors on the cat qubits. These excited pairs contribute to the bit-flip rate as soon as their energy-level spacing $\delta_n$ is significant with respect to $\kappa_{\text{conf}}$. For any $n$, the associated spacing $\delta_n$ converges to 0 when increasing the cat-qubit mean number of photons $|\alpha|^2$. This leads to a staircase pattern in the decreasing bit-flip error rate when increasing this mean number of photons. Such a behaviour was recently observed experimentally in a close concordance with theory~\cite{Frattini-2022}.  

\begin{figure}[t!]
    \centering
    \includegraphics[width=\columnwidth]{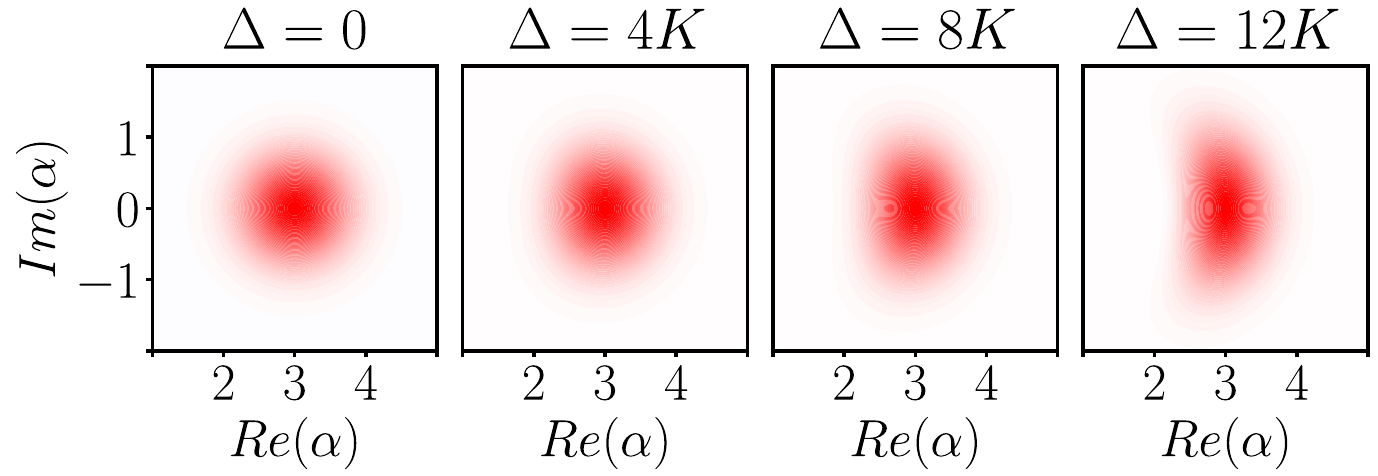}
    \vspace{-0.5cm}
    \caption{\label{fig:wigner}
    Wigner functions of the states $\ket{0}_{m}$ for $m=0,2,4,6$ and $\nbar = 9$. As $\Delta$ increases, the ground state of the detuned Kerr Hamiltonian is further distorted from a coherent state.}
\end{figure}

Assuming now that the squeezing drive is detuned with respect to the Kerr resonator frequency, the  Hamiltonian in the drives rotating frame becomes 
\begin{equation}
    \begin{split}
        \hat{H}_\Delta &= K\hat{a}^{\dagger 2}\hat{a}^{ 2} + \epsilon_2 \hat{a}^{\dagger 2} + \epsilon_2^* \hat{a}^{2} - \Delta  \adag \hata \\
        &= K(\hat{a}^{\dagger 2} - \alpha^{*2})(\hat{a}^{ 2} - \alpha^2) - \Delta  \adag \hata
    \label{eq:DKerr}
    \end{split}
\end{equation}
where $\Delta$ is the detuning of the two-photon drive. Importantly, for $\Delta = 2mK$ with $m \in \mathbb{N}$ a positive integer, this Hamiltonian admits two degenerate ground states. Note that these functioning points correspond to the non-dissipative limit of the bistable regimes discussed in~\cite{Roberts2019} ($(r_1,r_2)=(m,2m)$ in Fig. 3). We denote the ground states of even photon-number parity by $\ket{\mathcal{C}_\alpha^+}_m$ or $\ket{+}_m$ and odd parity by $\ket{\mathcal{C}_\alpha^-}_m$ or $\ket{-}_m$, and define
\begin{equation}
    \begin{split}
        &\ket{0}_m=\frac{1}{\sqrt{2}}(\ket{\mathcal{C}_\alpha^+}_m+\ket{\mathcal{C}_\alpha^-}_m) \\
        &\ket{1}_m=\frac{1}{\sqrt{2}}(\ket{\mathcal{C}_\alpha^+}_m-\ket{\mathcal{C}_\alpha^-}_m).
        \label{newstates}
    \end{split}
\end{equation} 
Note that for $m \geq 1$, $\ket{0}_m$ and $\ket{1}_m$ are no longer coherent states, but they remain located on distinct parts of the phase plane. Figure~\ref{fig:wigner} shows the Wigner representation of $\ket{0}_m$ for increasing values of the detuning $\Delta$.

Figure~\ref{fig:arche}(b) shows the eigenenergies of the Hamiltonian of Eq.~\eqref{eq:DKerr} for $m=3$. The Hamiltonian admits $m +1$  pairs of perfectly degenerate energy levels. We have grouped the higher eigenstates by their photon-number parity and denote them $\ket{\phi_n^\pm}_m$ with $e^\pm_{n,m}$ their respective energy. For $n \le m$, we denote by $\ket{\psi^0_n}_m$ and $\ket{\psi^1_n}_m$ the corresponding degenerate states located on the right- and left-hand side of phase space. They read
\begin{equation}
    \begin{split}
        &\ket{\psi^0_n}_m =\frac{1}{\sqrt{2}}\left(\ket{{\phi}_n^+}_m+\ket{{\phi}_n^-}_m\right) \\
        &\ket{\psi^1_n}_m =\frac{1}{\sqrt{2}}\left(\ket{{\phi}_n^+}_m-\ket{{\phi}_n^-}_m\right).
    \end{split}
\end{equation} 

Figure~\ref{fig:arche}(c) shows the energy level spacing $\delta_n$ between $\ket{{\phi}_n^+}$ and $\ket{{\phi}_n^-}$, as a function of the detuning. The combination of two mechanisms are thus demonstrated. As $\Delta$ increases, the mean number of photons in the cat states increases also because the frequency of the two-photon drive approaches the resonant frequency of higher transitions in the Kerr non-linear oscillator. Thus the energy level spacing $\delta_n$ between $\ket{{\phi}_n^+}$ and $\ket{{\phi}_n^-}$ diminishes as it does when the strength of the two-photon drive $|\epsilon_2|$ increases. For $\Delta = 2mK$ with $m \in \mathbb{N}$, the two-photon drive frequency becomes resonant with the transition $\ket{m}$ to $\ket{m+1}$ of the Kerr non-linear oscillator. As shown by the peaks around $\Delta~=~2mK$, new bistability points appear that include more and more pairs of degenerate eigenstates as $\Delta$ increases. We can also note that the energy gap increases between the first and second pair of degenerate eigenstates when $\Delta$ increases, as it does when $|\epsilon_2|$ increases. For the rest of this paper, we will use $|\alpha|^2 = |\epsilon_2/K|$ and $\bar{n} = \expval{a^\dagger a}$. Note that $\bar{n} \ne |\alpha|^2$ for $\Delta >0$ but rather $\bar{n} \approx |\epsilon_2/K|+\Delta/(2K)$ because the detuning increases the number of photons in the cavity. As it will become more clear in the next section, and according to Eq.~\eqref{eq:formuleRonan}, this is of particular interest to have multiple degenerate states in the spectrum, since the first $m +1$ pairs of eigenstates will not contribute to bit-flip errors.

We conclude this section by noting that for the particular choices of detuning $\Delta=2mK$, the $m +1$ pairs of degenerate eigenstates can be calculated analytically by diagonalizing two matrices of dimension $m+1$. In what follows, we assume that $\alpha$ is real. The first step is to displace the detuned Kerr Hamiltonian by~$\pm \alpha$
\begin{equation*}
    \begin{split}
        \tilde{H}_\pm &= \hat D(\pm \alpha) \hat{H}_\Delta \hat D(\mp \alpha) \\
         &= K \left( \adagsq \mp 2 \alpha \adag \right) \left( \hata^2 \mp 2 \alpha \hata \right) -  \Delta (\adag \mp\alpha)( \hata \mp \alpha)
    \end{split}
\end{equation*}
where $\hat D$ denotes the displacement operator. Evaluating this displaced frame Hamiltonian on the $n$-th Fock state $\ket{n}$ yields
\begin{equation}
    \begin{split}
        \tilde{H}_\pm \ket{n} & = \left[K (n-1) + 4 K |\alpha|^2 - \Delta \right] n \ket{n} \\ 
        & \mp \left[2K n - \Delta \right] \alpha \sqrt{n+1} \ket{n+1} \\
        & \mp \left[2K (n-1)  - \Delta \right] \alpha \sqrt{n} \ket{n-1}
    \end{split}
\end{equation}
For $\Delta = 2mK$, the Hamiltonians $\tilde H_\pm$ map the finite-dimensional Hilbert space spanned by the first $m+1$ Fock states to itself. This photon blockade was observed through a similar calculation in~\cite{lingenfelter2021} and used for preparing Fock states with weak Kerr strengths. In other words, these Hamiltonians are block diagonals. Therefore, by diagonalizing the associated block matrices of dimension $(m+1)\times(m+1)$, we can calculate the $(2m+2)$ first eigenstates of the Hamiltonian $\hat H_\Delta$, as being their displacements by $\mp\alpha$.  For instance for $m=1$, this leads to
\begin{equation*}
    \begin{split}
        \ket{0}_{m=1} &= \text{cos}(\theta)\ket{\alpha} + \text{sin}(\theta) \ket{\alpha,1} \\
        \ket{{1}}_{m=1} &= \text{cos}(\theta)\ket{-\alpha} - \text{sin}(\theta) \ket{-\alpha,1} \\
        \ket{\psi^0_{1}}_{m=1} &= \text{sin}(\theta)\ket{\alpha} - \text{cos}(\theta) \ket{\alpha,1} \\
        \ket{\psi^1_{1}}_{m=1} &= \text{sin}(\theta)\ket{-\alpha} + \text{cos}(\theta) \ket{-\alpha,1} \\
    \end{split}
\end{equation*}
with
\begin{equation*}
\theta=\arctan\left(\frac{2|\alpha|}{2|\alpha|^2-1+\sqrt{4|\alpha|^4+1}}\right).
\end{equation*}
Furthermore the eigenenergies corresponding to each pair are perfectly degenerate and the energy gap between the ground and first excited subspaces is given by
\begin{equation}
    \begin{split}
        e^\pm_{1,1}-e^\pm_{0,1}  &= 2 K \sqrt{1+4|\alpha|^4}.
    \end{split}
\end{equation}
 

\section{\label{sec:level4}Phase space confinement}

In the previous section, we argued that the slow suppression of bit-flip errors in resonant Kerr cat qubits can be explained through the non-degeneracy of excited energy-level pairs in the driven Kerr Hamiltonian. We also saw that, for specific choices of detuning $\Delta \in 2K\mathbb{N}$, the detuned Kerr Hamiltonian remarkably admits $m+1$ pairs of perfectly degenerate eigenstates. One can guess that this change in the spectrum should drastically improve the phase space confinement, and therefore lead to a faster suppression of induced bit-flip errors. To quantify this phase space confinement, we look at the average value of the observable $\hat S=\text{sign}(\hat X)=\text{sign}(\hat a+\hat a^\dag)$. This observable is very close to the two-photon dissipation conserved quantity $J_X=J_{+-}+J_{-+}$ defined in~\cite{Mirrahimi2014}, and quantifies whether the state of the oscillator state is located in the right or left half plane of phase space.  

The simulations were performed as follows. We initialize the system in the state $\ket{0}{\bra{0}}_m$ and let it evolve under the master equation 
\begin{equation}
    \dot{\hat\rho} = -i [\hat H_\Delta,\hat \rho] + \kappa_- \mathcal{\hat D}[\hata]\hat \rho + \kappa_+ \mathcal{\hat D}[\adag]\hat \rho + \kappa_\phi \mathcal{\hat D}[\adag \hata]\hat \rho
\end{equation}
where $\kappa_- = \kappa_1 (1 + n_{\text{th}})$, $\kappa_+ = \kappa_1 n_{\text{th}}$, $\kappa_1$ is the single-photon loss rate, $n_{\text{th}}$ is the thermal population and $\kappa_\phi$ is the dephasing rate. The excursions in phase space are quantified by $\langle \hat S \rangle_t = \Tr\bigl[\hat S\hat\rho(t)\bigr]$. By simulating the system over a long time of order $T=100/\kappa_1$, we can fit the above quantity $\langle \hat S\rangle_t$ to a single exponential $\exp(-2\Gamma_S t)$. For a strongly confined system, the above excursion rate $\Gamma_S$ will be very small.  The numerical simulations were performed in the basis of Kerr eigenstates with a truncation to 20 eigenstates. The
simulation code was written using the QuTip
package~\cite{johansson2012}.

\begin{figure}[t!]
    \centering
    \includegraphics[width=\columnwidth]{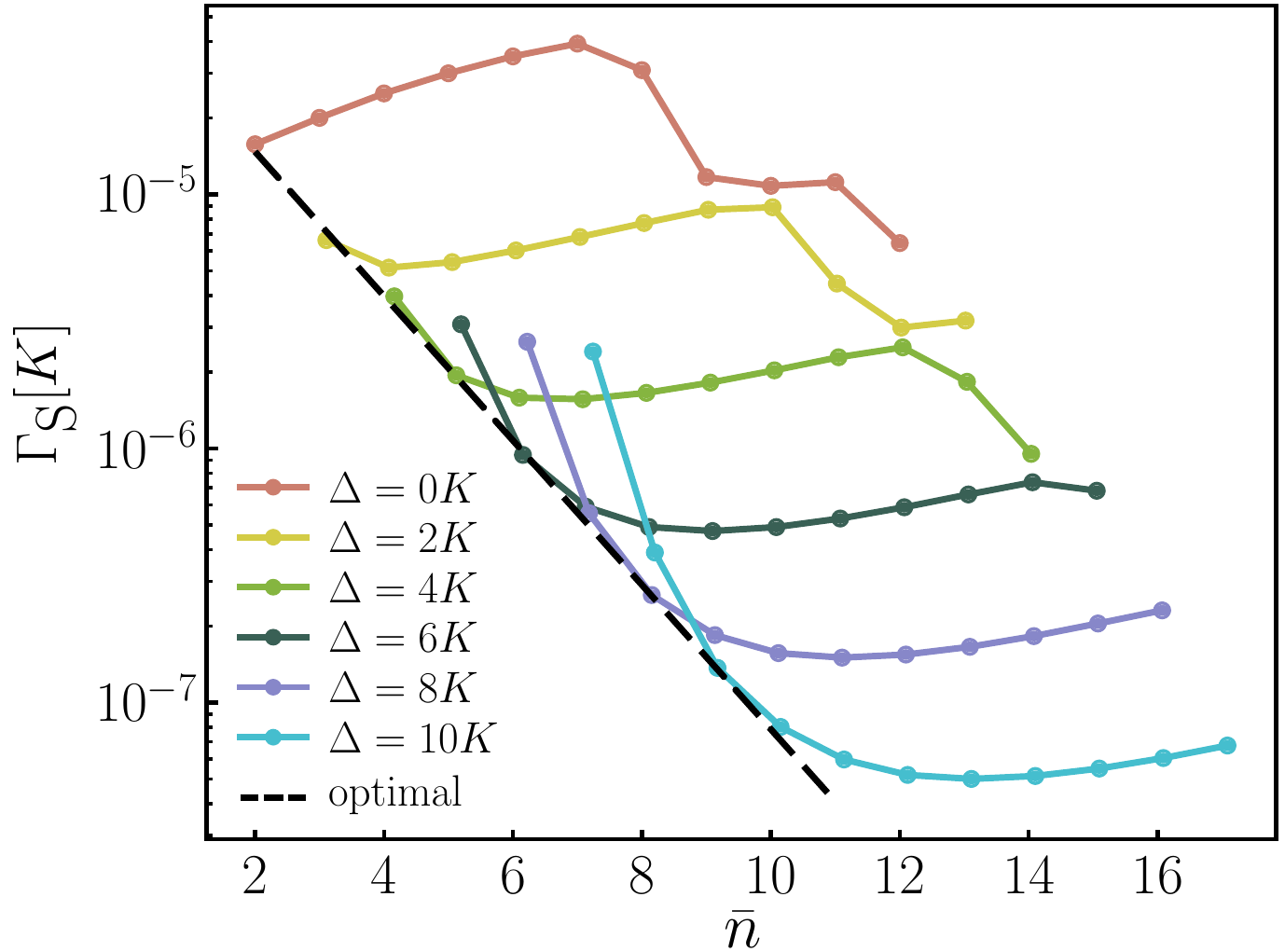}
    \vspace{-0.5cm}
    \caption{\label{fig:bitflip}
         Phase space excursion rate $\Gamma_\text{S}$, \ie~rate at which the expectation value $\langle \hat S \rangle$ of the confinement observable $\hat S=\textrm{sign}(\hata+\adag)$ decays to zero. This rate is plotted as a function of $\bar{n} = \langle\adag \hata\rangle$ for $\kappa_1 = 10^{-3}K$, $n_{\text{th}}=10^{-2}$ and $\kappa_\phi = 10^{-5}K$ and  for different detuning values. As $\Delta=2mK$ increases, more and more eigenstates come in degenerate pairs and do not contribute to such phase space excursions. Thus the rate $\Gamma_S$ decreases and saturates at a lower value when increasing $\bar n$. For a fixed value of $\bar n$, however, there is an optimal choice of $\Delta$ that minimizes the rate $\Gamma_S$: $\Delta_{\text{opt}} \approx \nbar K$. The associated optimal rate $\Gamma^{\text{opt}}_S$ scales as $e^{-\gamma\bar n}$, with $\gamma\approx 0.65$ (note that this value corresponds to the specific values of $\kappa_1,n_{\text{th}} \text{ and } \kappa_\phi$ chosen).
    }
\end{figure}

Figure~\ref{fig:bitflip} presents such simulations for $\kappa_1 = 10^{-3}K,n_{\text{th}} = 10^{-2} \text{ and } \kappa_\phi = 10^{-5}K$. The rate $\Gamma_S$ is evaluated as a function of $\bar{n} = \expval{\adag \hata}$. We remind that $\bar{n} > |\epsilon_2/K|$ for $\Delta>0$. As expected, we benefit from a much stronger phase space confinement for $\Delta = 2mK$, $m \ge 1$ compared to $\Delta = 0$. Note however that for a fixed low $\bar{n}$, the phase space excursions can be faster when increasing $\Delta$. This is mainly due to the fact that for large $\Delta$, the two-photon drive amplitude $|\epsilon_2|$ becomes too weak to separate the states $\ket{0}_m$ and $\ket{1}_m$ in phase space.  More precisely, for a fixed $\bar{n}$, an optimal $\Delta$ can be found to minimize the excursion rate $\Gamma_S$: $\Delta_{opt} \approx \nbar K$. Furthermore, this optimized rate $\Gamma_S$ scales as $e^{-\gamma\bar{n}}$ with $\gamma \approx 0.65$ (dashed line in Fig.~\ref{fig:bitflip}).

\begin{figure}[t!]
    \centering
    \includegraphics[width=\columnwidth]{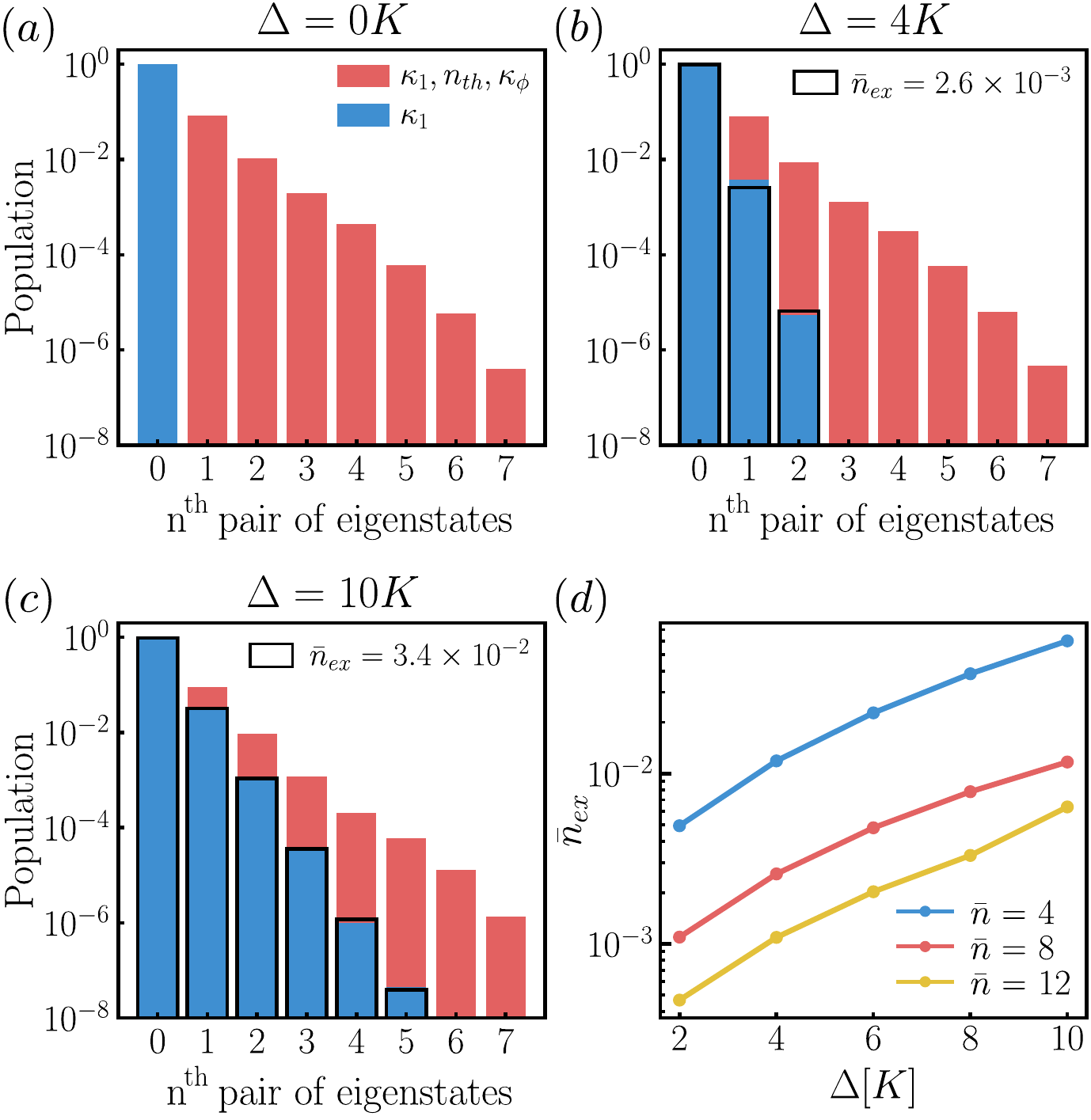}
    \vspace{-0.5cm}
    \caption{\label{fig:thermalstate}
    Steady state population of the eigenstates of the detuned Kerr Hamiltonian for (a) $\Delta = 0K$, (b) $\Delta = 4K$ and (c) $\Delta = 10K$, and for $\nbar = 10$, $\kappa_1 = 10^{-3}K$, $n_{\text{th}}=10^{-2}$ and $\kappa_\phi = 10^{-5}K$. Starting from $\ket{0}_m$, a time of the order of $1/\kappa_1$ is needed to reach this mixed steady state. In presence of thermal excitation or dephasing, all eigenstates are populated following an exponential distribution that does not significantly change with the detuning. However, under the absence of these processes, and thus under the sole effect of single-photon loss channel, only the degenerate pairs are populated. In the latter case, the populations of these degenerate eigenstates reach a distribution close to the Boltzmann distribution of a thermal state. (d) The effective mean number of excitations $\bar n_{ex}$ of the mixed steady state of the driven detuned Kerr under single-photon loss as a function of the detuning and for different values of $\nbar$. 
    }
\end{figure}

The Kerr cat, on resonance or detuned, undergoes an important amount of leakage due to thermal excitation and dephasing, as can be seen on Fig.~\ref{fig:thermalstate}(a), (b) and (c) with the red bar charts. The situation is however different when the system is only subject to  single-photon loss. In this case, the resonant Kerr cat has a very small leakage because the coherent states are not affected by the annihilation operator $\hata$. However the detuned Kerr cat states are no longer perfect eigenstates of  $\hata$. Thus, the single-photon loss induces significant leakage to the other energy levels. 

On Fig.~\ref{fig:thermalstate}(a), (b) and (c), the blue bar charts represent the population on the $n$-th pair of eigenstates when the detuned Kerr cat is only undergoing single-photon loss and has reached a steady state. As it can be seen in these charts, the single-photon loss only induces leakage to the eigenstates in the degenerate part of the spectrum. Indeed, for the detuned Kerr cats, because the manifold Span$\{{\ket{\psi^0_n}}_m\}_{n\in  [\![ 0,m]\!]}$ is equal to the manifold Span$\{{\ket{\alpha,n}}\}_{n\in  [\![ 0,m]\!]}$, which is stable under the application of $\hata$, single-photon loss only populates the degenerate pairs of eigenstates. Interestingly, as it can be seen in plots (b) and (c), this degenerate manifold reaches a mixed state with an  exponential distribution on degenerate pairs close to a thermal state distribution with an effective non-zero temperature.  Figure~\ref{fig:thermalstate}(d) shows the effective mean number of excitations $\bar n_{ex}$ in this degenerate manifold. In particular, we see that it increases very rapidly with $\Delta$. Note that this mean excitation number does not depend on $\kappa_1$. In order to work with well defined qubits for fault-tolerant quantum computation, this leakage needs to be suppressed. This can be done through the addition of dissipative processes refocusing the population to the ground manifold of the detuned Kerr Hamiltonian. 

\section{\label{sec:colored}Detuned Kerr cats with colored dissipation}

Bosonic qubits are often solely defined through their codespace, which is a two-dimensional subspace of an oscillator infinite-dimensional Hilbert space. However, a full definition of bosonic qubits should include a complete mapping from the oscillator space to this code space to characterize any leakage that may occur outside of it. Indeed, once readout of the bosonic qubit is performed, it is important to be able to associate any potential leakage out of codespace to one of the qubit computational states. For instance, in the case of GKP qubits~\cite{Gottesman2001,CampagneIbarcq2020}, the full oscillator space is mapped to the qubit through a grid-like separation of phase space. Any small displacement away from the codespace would thus be properly taken into account.

For cat qubits, such a mapping can be performed with dissipative stabilization by associating any initial state to its infinite-time steady state once reconverged to the cat-qubit codespace. As such, it is essential to have a process that eliminates leakage in order to rigorously define cat qubits. For the detuned Kerr cat qubit introduced in this paper, this dissipative stabilization cannot be realized with the driven two-photon dissipation $\mathcal{D}[\hat{a}^2 - \alpha^2]$ as the ground states of the Hamiltonian are not coherent states anymore, thus are not dark states of this dissipative super-operator. Instead, we show in this section that the stabilization of detuned Kerr cat qubits may be achieved with a colored dissipation.

The colored dissipation method introduced in Ref.~\cite{Putterman2022} consists in enabling precise frequency decays to bring back leakage to the ground eigenspace of the detuned Kerr Hamiltonian. The full system can be modelled as follows,
\begin{equation}
    \begin{split}
        \frac{d \hat \rho}{dt} &= -i[\hat H_\Delta + \hat H_\text{color},\hat \rho] + \kappa_1(1+n_{\text{th}}) \mathcal{\hat D}[\hat a]\hat \rho \\ 
        & + \kappa_1 n_{\text{th}} \mathcal{\hat D}[\adag]\hat \rho + \kappa_\phi \mathcal{\hat D}[\adag \hata]\hat \rho + \kappa_f \mathcal{\hat D}[\hat f_M]\hat \rho
        \label{eq:color}
    \end{split}
\end{equation}
where
\begin{equation}
        \hat H_\text{color} = g \hat a \hat f_1^\dagger e^{i\Delta_f t} + J \sum_{j=1}^{M-1}\hat f_j \hat f_{j+1}^\dagger + \hc
\end{equation}
is an interaction Hamiltonian between the Kerr cat mode $\hata$ and all filter modes $\hat f_1,...,\hat f_{M}$, with respective frequencies $\omega_a$ and $\omega_f$. Together with the dissipation of the last filter mode in $\mathcal{D}[\hat{f}_M]$, this Hamiltonian approximates an ideal band-pass filter centered on the  frequency $\omega_f=\omega_a+\Delta_f$, with half bandwidth $\kappa_f = 2J$ ~\cite{Putterman2022}.

The main point of this filter is to allow the relaxations $\ket{\phi_n^\pm}_m \rightarrow \ket{\phi_{n-1}^\mp}_m$ to occur, thus countering the leakage from the code space, while filtering out the transitions $\ket{\phi_n^\pm}_m \rightarrow \ket{\phi_n^\mp}_m$ that solely induce photon number parity jumps without reducing state leakage. Indeed, while the first kind of transition features an energy difference $\Delta_f \gtrsim 4K \nbar$, the second kind has a typical energy difference $\Delta_f \approx 0$, where the equality is verified within the exactly-degenerate subspace of the detuned Kerr Hamiltonian (see Fig.~\ref{fig:arche}(b)). Note in particular that all of these transitions swap the cat state parity, thus leakage elimination with colored dissipation is performed while inducing parity swaps. This is beneficial for certain gates, as it ensures the correction of first-order non-adiabatic phase-flip errors (see Ref.~\cite{Putterman2022} for details).

Similarly as in Ref.~\cite{Putterman2022}, we set the filter detuning frequency at the first excited to ground state transition frequency, \ie~$\Delta_f \approx 4K \bar{n}$. We also set $\kappa_f = 2J = \Delta_f / 5$ and $g = \kappa_f / 5$ like for the regular colored Kerr cat, such that adiabatic elimination of the ancillary filter modes can be performed. In the limit of a large number of filter modes, this colored dissipation leads to an engineered dissipation rate $\kappa_{1,\textrm{eng}}=4g^2/\kappa_f$. A notable difficulty in the design of this system is to filter out the $\Delta_f \approx 0$ transitions --- in order not to induce persistent phase-flip errors --- while maintaining a maximum number of higher energy transitions. As such, it may be relevant to feature multiple band-pass filters at various transition frequencies, so-called "colors".

\begin{figure}[t!]
    \includegraphics[width=\columnwidth]{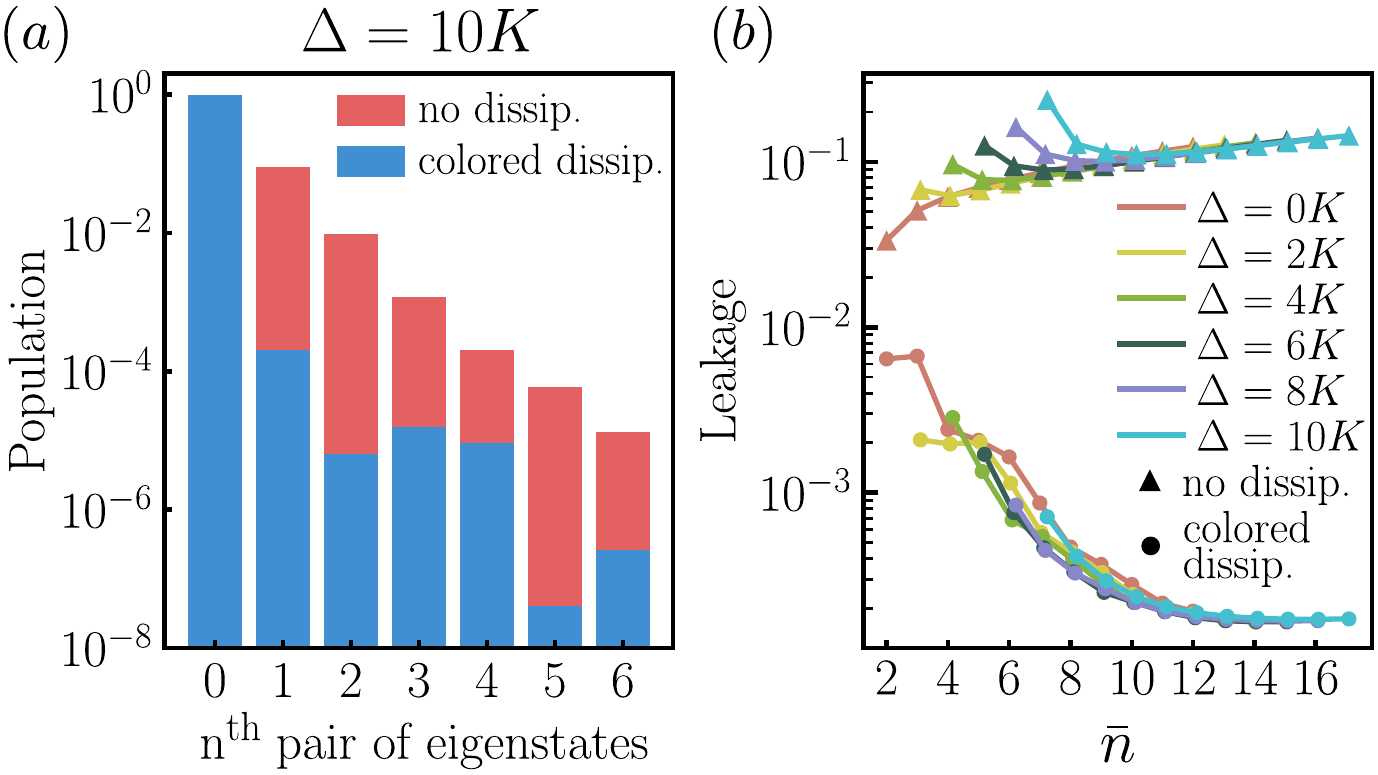}
    \caption{\label{fig:coloredthermalstate}
    (a) Thermal Population of the eigenstates of the detuned Kerr Hamiltonian for $\Delta = 10K$ for $\nbar = 10,\kappa_1 = 10^{-3}K, n_{\text{th}}=10^{-2},$ and $\kappa_\phi = 10^{-5}K$ without colored dissipation (in red) and with colored dissipation (in red). The colored dissipation reduces the leakage by nearly three orders of magnitudes. (b) Leakage out of the code space as a function of $\nbar$ and for different values of $\Delta$. The colored dissipation works regardless of the value of the detuning.
    }
\end{figure}

The population of the steady state of the detuned Kerr cat is shown in Fig.~\ref{fig:coloredthermalstate}(a), with (blue) or without (red) colored dissipation and for a fixed detuning $\Delta = 10K$, cat size $\nbar = 10$, in presence of several leakage-inducing dissipative terms such as thermal excitation and dephasing. Several orders of magnitude of reduction in the amount of leakage are therefore demonstrated once this colored dissipation is added. In particular, the rate of this colored dissipation, or the number of colors, can be adjusted to further reduce leakage. The leakage reduction induced by colored dissipation is shown in Fig.~\ref{fig:coloredthermalstate}(b) as a function of the mean number of photons $\bar{n}$ and for different values of the detuning $\Delta$.

\begin{figure}[t!]
    \includegraphics[width=\columnwidth]{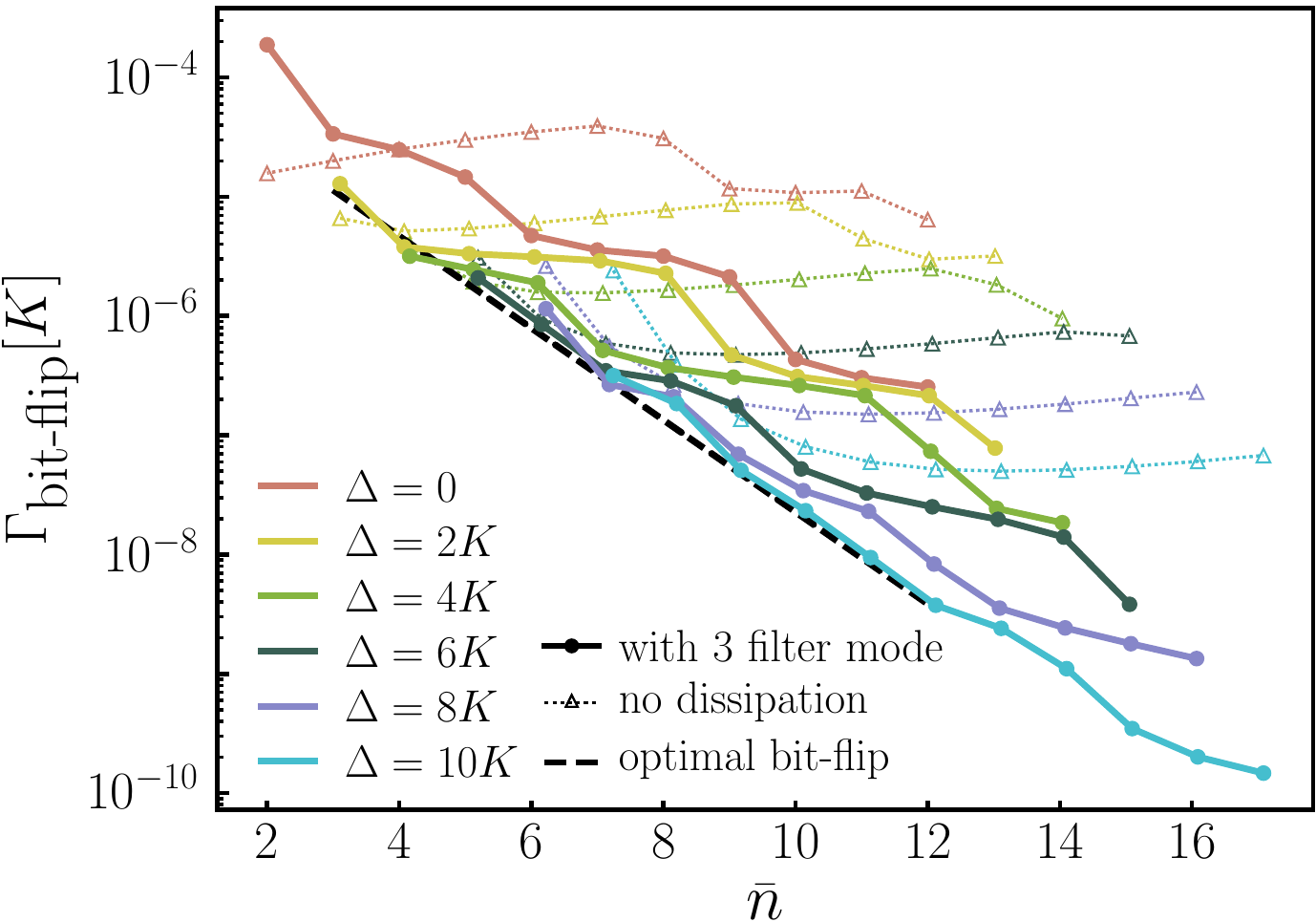}
    \vspace{-0.5cm}
    \caption{\label{fig:coloredbitflip} 
    Bit-flip error rate as a function of $\nbar$ for different values of the detuning $\Delta$ and for $\kappa_1 = 10^{-3}K, n_{\text{th}}=10^{-2}$ and $\kappa_\phi = 10^{-5}K$. The transparent curves represent the phase space excursion rate without colored dissipation (same as Fig.~\ref{fig:bitflip}), and the plain curves represent the bit-flip rate with colored dissipation. For a fixed value of $\bar n$, there is an optimal choice of $\Delta$ that minimizes the rate $\Gamma_\text{bit-flip}$. The associated optimal rate $\Gamma^{\text{opt}}_\text{bit-flip}$ scales as $e^{-\gamma\bar n}$, with $\gamma\approx 0.89$ (note that this value corresponds to the specific values of $\kappa_1,n_{\text{th}} \text{ and } \kappa_\phi$ chosen).
    }
\end{figure}

Once a dissipative scheme stabilizing the two-dimensional subspace (preventing state leakage to accumulate) is precised, we may speak of a well defined qubit. The logical Pauli operators of the cat qubit are defined as $Z=\text{sign}(\hata+\adag)$ and $X=\exp(i\pi\adag\hata)$. Note that, by reducing leakage, the colored dissipation also reduces the bit-flip rate. This can be seen from the rate of phase space excursions $\Gamma_S$ that now corresponds to the bit-flip error rate $\Gamma_\text{bit-flip}$. Similar simulations to those performed in the previous Section, but now in the presence of colored dissipation, are provided in Fig.~\ref{fig:coloredbitflip}. In these simulations, we let the system evolve from the state $\ket{0}_m$ under the master equation~\eqref{eq:color} for a time of order $10/\kappa_1$, and fit $\expval{Z}_t=\langle \hat S \rangle_t$ to an exponential $\exp(-2\Gamma_{\text{bit-flip}}t)$. Figure~\ref{fig:coloredbitflip}(a) shows the resulting bit-flip error rates for different values of $\bar{n}$ and detuning, with $\kappa_1 = 10^{-3}K$, $n_{\text{th}} = 10^{-2}$ and $\kappa_\phi = 10^{-5}K$. The simulations were performed in the Kerr eigenstate basis with a truncation of 14 eigenstates. Following Ref.~\cite{Putterman2022}, we assume that there is at most one photon in all $N$ filter modes, such that they can all be simulated at once using a Hilbert space of dimension $N+1$.

Note that for a small number of filter modes, the addition of colored dissipation can induce additional phase-flip errors. Thus, in practice, a certain number of modes are needed so that the additional phase-flip errors become negligible compared to those caused by the intrinsic single-photon loss. For the parameters in our simulations, this level of filtering is reached using three modes. In practice, one might want to reduce the engineered colored dissipation rate $\kappa_{1,eng}$, in order to avoid further phase-flip errors due to the absence of a perfect cut-off, or to limit the non-adiabatic gate errors introduced in the next section. Increasing the detuning enables to reach an equivalent bit-flip error rate, but for a weaker colored dissipation rate.

\section{\label{sec:level6}Bias-preserving gates}

For a universal set of logical operations, it is required to be able to prepare cat states in both $\ket{0}_c$ and $\ket{+}_c$~\cite{Guillaud2019}. The preparation of $\ket{+}_c$ can be achieved by adiabatically increasing both the two-photon drive strength and detuning. The best adiabatic path can be optimized in a similar way as Ref.~\cite{Yanagimoto2019} leading to a preparation analogous to on-resonance Kerr cats. The preparation of $\ket{0}_m$ can be achieved similarly, starting from a coherent state stabilized with an on-resonance Kerr Hamiltonian, and adiabatically increasing the detuning. During this preparation, the rate of bit-flip errors will be equivalent to the on-resonance Kerr, and will then recover the rate described in Fig.~\ref{fig:coloredbitflip}.

The implementation of bias-preserving gates on cat qubits, as required for a universal set of fault-tolerant logical gates, are achieved using two types of dynamics, so-called Zeno and topological gates~\cite{Mirrahimi2014,Guillaud2019,Puri2020}. A Zeno gate typically makes use of the quantum Zeno effect to perform a rotation of an arbitrary angle $\theta$ around the Z axis by applying a weak near-resonant drive. In the rotating frame of the two-photon drive, such a Zeno gate can be modelled by the addition of the Hamiltonian 
\begin{equation}
    \hat H_Z(t) = \epsilon_Z(t) (\hat a^\dagger + \hat a)
    \label{eq:Z}
\end{equation}
to Eq.~\eqref{eq:DKerr}. Here, $\epsilon_Z(t)$ represents a slowly varying modulation of the driving field amplitude. A rotation around the Z axis of the Bloch sphere is then realized with an angle $4 \sqrt{\nbar} \int_0^T \epsilon_Z(t) dt$, where T is the duration of the gate. Accelerating such a gate usually comes at the expense of additional leakage out of the cat-qubit subspace due to non-adiabatic effects. In the absence of decoherence, and by benefiting from the adiabatic theorem with exponential accuracy~\cite{teufel:book}, it is possible to engineer pulses $\epsilon_Z(t)$ where these higher-order effects are carefully suppressed, thus reaching extremely fast gates~\cite{Xu2021}. Two-qubit entangling gates (rotations around $ZZ$ axis) can be performed in a similar manner~\cite{Mirrahimi2014}. These gates rely on the protection provided by the Hamiltonian gap and thus straightforwardly used in the context of detuned Kerr cats, while preserving the better scaling of bit-flip type errors.

\begin{figure}[t!]
    \includegraphics[width=\columnwidth]{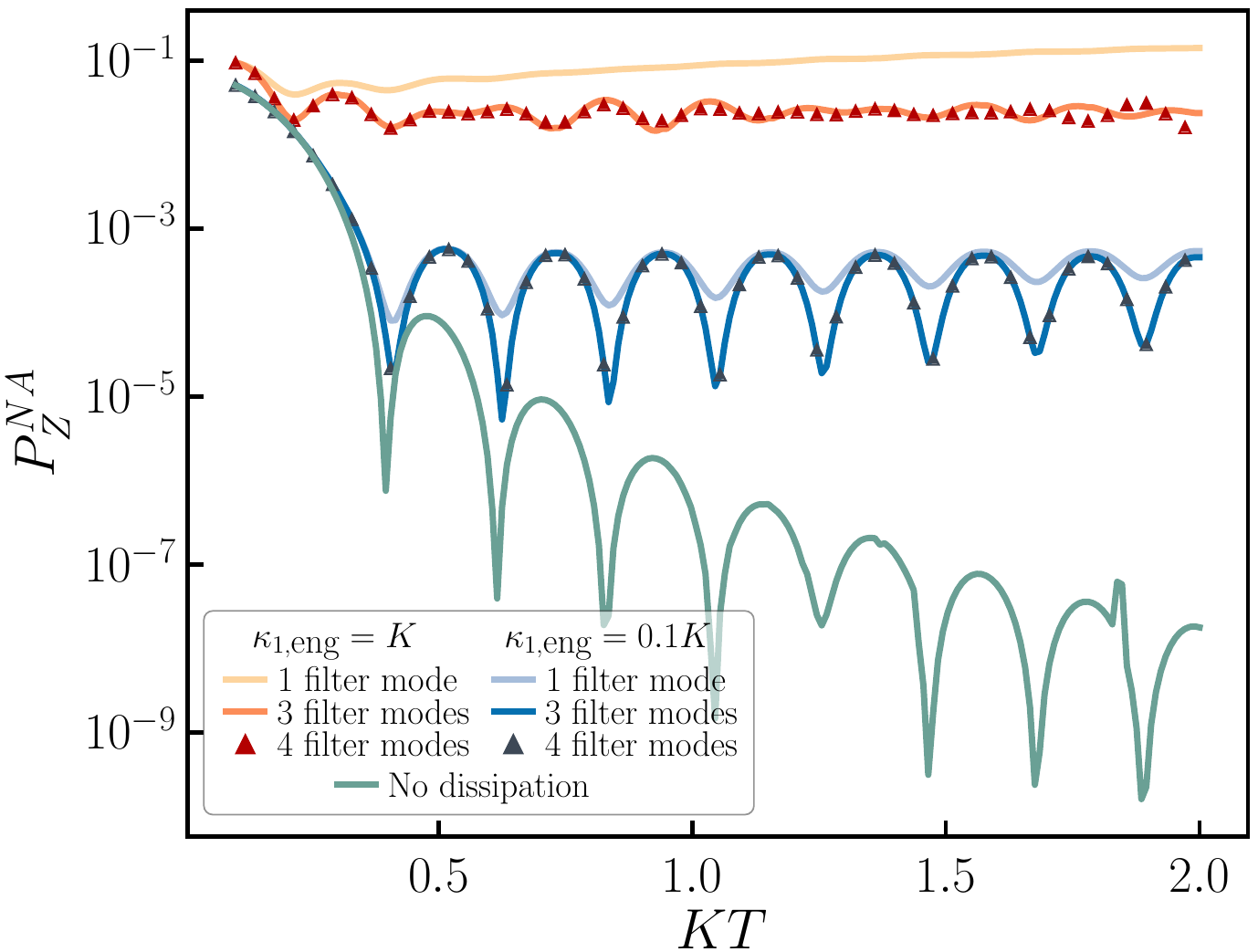}
    \vspace{-0.6cm}
    \caption{\label{fig:Zgate}
    Non-adiabatic error probability $p_Z^{NA}$ for a Zeno $\pi$-rotation around the $Z$ axis, realized with a Gaussian pulse $\epsilon_Z(t)$ for $\nbar = 8$. Here $T$ is the gate duration and is expressed in units of $1/K$. Without dissipation, the non-adiabatic error probability can be reduced to as low as $10^{-8}$ for gate times of order $1/K$. But, with colored dissipation of strength $\kappa_{1,eng} = K$ such non-adiabatic error probability saturates above $10^{-2}$. Dividing by 10 the colored dissipation amplitude reduces the non-adiabatic error probability by approximately two orders of magnitudes. $p_Z^{NA}$ is also plotted for different number of filter modes in the colored dissipation, showing that it is not necessary to consider more than 3 filter modes here.
    }
\end{figure}

We can distinguish two sources of phase-flip errors when applying such a Zeno gate. The first one corresponds to those induced by the undesired single-photon loss, with probability $p_Z = \nbar \kappa_1 T$. The second one is non-adiabatic errors, with probability $p_Z^{NA}$, which are created because the gate induces leakage out of the code space, and consequently this leakage is mapped to code space  errors after being removed by the engineered dissipation process. Without dissipation, the non-adiabatic errors can be exponentially suppressed with the gate duration, using the super-adiabatic pulse designs~\cite{Xu2021}. However, when combined with colored dissipation, the Z rotation induces further non-adiabatic errors. The detuned Kerr cats suffer from these non-adiabatic errors just as the resonant ones. However, for the same bit-flip error rate, the strength of the colored dissipation of the detuned Kerr cats can be reduced compared to the resonant ones, leading therefore to smaller non-adiabatic error rates. For instance, as can be seen on Fig.~\ref{fig:coloredbitflip}, at $\nbar = 8$ and for $\Delta = 10K$, the bit-flip rate of the detuned Kerr cats (even without colored dissipation) is smaller than the bit-flip rate of the resonant one with colored dissipation. Note however that the colored dissipation is still needed to counter leakage, and to work with well defined qubits. Figure~\ref{fig:Zgate} shows the reduction in non-adiabatic errors on a Z gate when reducing the strength of the colored dissipation from $\kappa_{1,eng}=K$ to $\kappa_{1,eng}=K/10$.

The topological gates are based on a deformation of the code space. Typically,  a Pauli X gate is performed through a $\pi$-rotation of the phase space by rotating the two states confined by the Kerr dynamics $\ket{\alpha(t)} = \ket{\alpha e^{i \theta(t)}}$ (with $\theta(T)=\pi$), while applying a so-called feed-forward Hamiltonian ${\hat H(t)=-\dot\theta(t) \hat a^\dagger \hat a}$. Here, nothing in the dynamics is specific to having coherent states as confined states, and therefore the Pauli X gate can be implemented in the exact same manner for the detuned Kerr cats. Less trivially, the CNOT gate introduced in~\cite{Puri2020} can also be adapted to the detuned Kerr cats. It consists in rotating the confinement of the target qubit conditionally on the state of the control qubit as well as applying a CNOT feed-forward Hamiltonian
\begin{equation}
    \begin{split}
        &\hat H(t) = K(\adagsq_c - \alpha_c^2)(\hata^2_c - \alpha_c^2) - \Delta \adag_c \hata_c - \Delta \adag_t \hata_t \\
        &+ K\left[\adagsq_t - \alpha^2_t e^{-2 i \theta(t)} \left(\frac{\tilde{\alpha}_c - \adag_c}{2\tilde \alpha_c}\right) - \alpha_t^2 \left(\frac{\tilde{\alpha}_c + \adag_c}{2\tilde\alpha_c}\right) \right] \\ 
        &\times \left[\hata^2_t  - \alpha^2_t e^{-2 i \theta(t)} \left(\frac{\tilde{\alpha}_c - \hata_c}{2\tilde\alpha_c}\right) - \alpha_t^2 \left(\frac{\tilde{\alpha}_c + \hata_c}{2\tilde\alpha_c}\right) \right] \\
        &+ \dot\theta(t) \frac{2 \tilde{\alpha}_c - \adag_c - \hata_c}{4 \tilde\alpha_c} \otimes (\adag_t \hata_t - \nbar_t) 
        \label{eq:cnot-drive}
    \end{split}
\end{equation}
with
\begin{equation}
        \alpha = \sqrt{\varepsilon_2/K}, \quad
        \tilde{\alpha} = {\bra{0}}_m(\adag+\hata)\ket{0}_m/2,
\end{equation}
where $\Delta=2mK$, $\theta(T) = \pi$ and $\hata_c$ and $\hata_t$ are the annihilation operators of the control and target modes respectively. 

\section{\label{sec:level9}Conclusion}

On the path towards a fault-tolerant quantum processor, biased-noise qubits are promising candidates as they significantly reduce the required overhead for error correction. By focusing on the case of Kerr cat qubits~\cite{Puri2017,Grimm2020,Frattini-2022}, we show in this paper that an extra control parameter, the detuning of the the two-photon drive with respect to the Kerr oscillator's resonance, plays a central role in the structure of the spectrum of the confinement Hamiltonian. For particular choices of detuning given by even multiples of the Kerr strength, not only the two encoding qubit states are perfectly degenerate, but also the excited states come in perfectly degenerate pairs. This strong symmetry significantly contributes to the properties of the encoded qubits. By keeping the information well confined in the left and right half planes of the phase space, such a degenerate spectrum strongly suppresses the phase space excursions induced by leakage mechanisms such as photon loss, thermal excitation, or photon dephasing. This Hamiltonian confinement can then be safely combined with a colored dissipation process countering the leakage and therefore leading to well-defined qubits. The degeneracy of the spectrum ensures that even a weak engineered dissipation is enough to benefit from a strong suppression of bit-flip errors. The weakness of the required dissipation provides room for engineering fast high-fidelity bias-preserving gates, where the information is not lost through the engineered dissipation channel. 

The analysis presented in this work is based on the effective static Hamiltonian given in equation~(\ref{eq:DKerr}). One may legitimately wonder whether this model is reasonable or not. Indeed, within this effective theory, it would seem that arbitrarily increasing the detuning of the squeezing drive would result in an arbitrary number of pairs of excited states being perfectly degenerate. In practice however, a number of approximations have been made to obtain this effective dynamics; and it would be interesting to investigate how higher order terms in the rotating wave approximation (RWA) affect the benefits of the perfect degeneracy of the eigenstates, or to derive a more general theory including beyond RWA effects, following the recent works of Refs.~\cite{Venkatraman2022}. However, even though the theory in this paper does not include the study of such effects, during the writing of this paper we have been informed that the experimental observations by our colleagues at Yale University~\cite{RodrigoExperiment} confirm the improvement of phase-space confinement property for specific detuning values mentioned in our paper.

\begin{acknowledgments}
D.R would like to thank Yiwen Chu for her supervision on this project. All the authors would like to thank Rodrigo Cortiñas, Raphaël Lescanne, Paul Magnard and Nathanaël Cottet for many enlightening conversations about the possible experimental realizations of this proposal. We acknowledge funding from the Plan France 2030 through the project ANR-22-PETQ-0006.
\end{acknowledgments}

\bibliography{biblio}

\end{document}